\documentclass[12pt,preprint]{aastex}

\shorttitle{On the heating of source of the Orion hot core.}
\shortauthors{De Vicente, Mart\'\i n-Pintado, Neri and Rodr\'\i guez-Franco}
\begin{document}
\title{ On the heating source of the Orion KL hot core}

\author{P. de Vicente, J. Mart\'\i n-Pintado\altaffilmark{1}}
\affil{ Observatorio Astron\'omico Nacional, Apartado 1143,
28080 Alcal\'a de Henares, Spain}
\email{vicente@oan.es, martin@oan.es}

\author{R. Neri}
\affil{Institut de Radioastronomie Millimetrique, Rue de la Piscine,
St Martin de Heres, France}
\and

\author{A. Rodr\'\i guez-Franco}
\affil{ E.U. de Optica, Dpto. de Matem\'atica Aplicada, UAM, Madrid, Spain}

\altaffiltext{1}{CSIC, IEM, Dpto. F\'\i sica Molecular, Serrano 123, 28006 Madrid, Spain}

\begin{abstract}
   We present images of the J=10--9 rotational lines of HC$_3$N in the
vibrationally excited levels 1v7, 1v6 and 1v5 of the hot core (HC) in 
Orion KL. The images show that the spatial distribution and the size 
emission from the 1v7 and 1v5 levels are different. While the J=10--9 1v7 line 
has a size of $4''\times 6''$ and peaks $1.1''$ NE of the 
3 mm continuum peak, the J=10--9 1v5 line emission is 
unresolved ($<3''$) and peaks $1.3''$ south of the 3 mm peak. This is a 
clear indication that the HC is composed of condensations with 
very different temperatures (170 K for the 1v7 peak and $>230$ K for the 1v5 
peak). The temperature derived from the 1v7 and 1v5 lines increases
with the projected distance to the suspected main heating source 
I. Projection effects along the line of sight could explain the
temperature gradient as produced by source I. However, the
large luminosity required for source I, $>5\, 10^{5}$ L$_\odot$, 
to explain the 1v5 line suggests that external heating by this source 
may not dominate the heating of the HC. Simple model calculations of 
the vibrationally excited emission indicate that the HC can be 
internally heated by a source with a luminosity of $10^5$ L$_\odot$,
located $1.2''$ SW of the 1v5 line peak ($1.8''$ south of source I).
We also report the first detection of high-velocity gas from
vibrationally excited HC$_3$N emission. 
Based on excitation arguments we conclude that the main heating source 
is also driving the molecular outflow. We speculate that all the 
data presented in this letter and the IR images are consistent with 
a young massive protostar embedded in an edge-on disk.
\end{abstract}


\section{Introduction}

Located at only 450 pc from the sun, the Orion KL region has been 
the prototype source to study the formation of high mass stars. 
This region shows all the signatures of massive 
star formation. It contains a plethora of IR cores 
($\simeq 20$) \citep{dougados93,gezari98}, H$_2$O masers 
\citep{genzel81,gaume98}, millimeter continuum emission 
\citep{plambeck95,blake96}, compact radio sources 
\citep{churchwell87,menten95}, and molecular hot cores and 
fast molecular outflows \citep{blake96,wright96,chandler97,rodriguez99b}.

  It is believed that most of the IR sources in the Orion KL  
nebula are not self luminous but show reprocessed emission escaping 
through inhomegeneities of the dense material surrounding a deeply embedded
source. Recently, \citet{menten95} and
\citet{gezari98} proposed that the main heating source in the region is 
radio source I.  
The lack of an IR 
counterpart to source I has led some authors to invoke large intrinsic 
foreground extinction towards source I. However, \citet{blake96},\citet{wright96} and 
\cite{chandler97} have shown that 
the dust and molecular peak emission, coincide neither
with radio source I nor with IRc2. The main peak of continuum emission at
1.3 mm, is $1''$ east from source I. Furthermore, based on the distribution of the HC$_3$N
J=24--23 line in the 1v7 vibrationally excited state, \citet{blake96}
have proposed that source I is the dominant energy source in the region, 
and that there is no evidence of internal heating within the molecular 
hot core (HC). However, \citet{kaufman98} based on a model for the heating
of the HC concluded that it is internally heated by young 
embedded stars.

  In this letter we investigate the thermal structure of the Orion 
KL HC by 
using the J=10--9 rotational line of HC$_3$N in the vibrational excited 
levels 1v7, 1v6 and 1v5. If the hot core material is not internally 
heated and source I is the main heating source one would expect the 
emission from the HC$_3$N rotational lines in the vibrational 
states (hereafter HC$_3$N*) with larger excitation energy to peak 
close to source I. We find that 
the emission of the HC$_3$N J=10--9 lines in the different vibrational 
levels peak at different positions indicating that the HC is internally heated. 

\section{Observations}

We performed 3 snapshots of the Orion KL HC with
three antennas of the IRAM Plateau de Bure Interferometer (PdBI)
for the following hour angle ranges: (-2.9,1.8), (-3.4,-1.0) and (1.1,2.3). 
The instrumental parameters for these observations have been described by 
\citet{devicente00}. These observations were made 
with the C1 and Dc configurations and the 500 MHz 
IF band was centered 
at 91.2 GHz in order to measure simultaneously the HC$_3$N J=10--9,1v5,1f, 
J=10--9,1v6,1e, J=10--9,1v7,1e and J=10--9,1v7,1f transitions. The half power 
size of the synthesized beam was 6.6$''$$\times$3.8$''$ PA: 2$^o$. 
We used the sources 3C111 (3.8 Jy) and 0458-020 (1.8 Jy) as flux 
density and phase calibrators and 3C454.3 as bandpass calibrator.

\section{Results}
\begin{figure}
\plotone{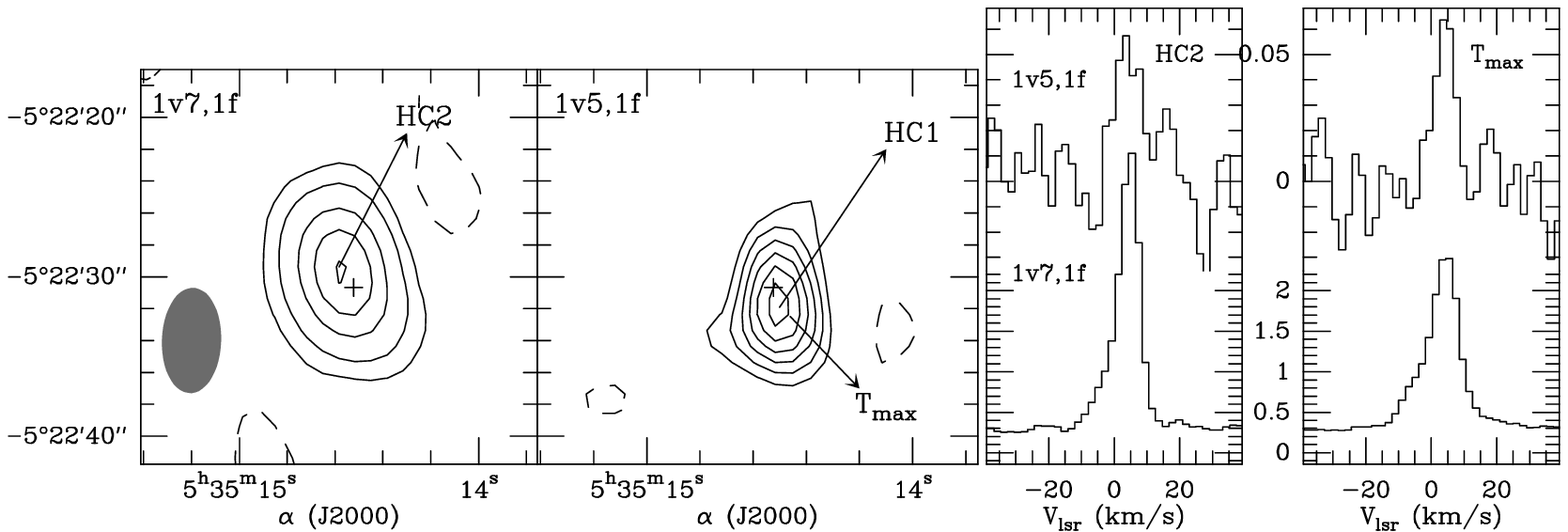}
\caption{Integrated intensity map of the main gaussian line 
(velocity interval from 3.5 to 7.7 kms$^{-1}$) for the 1v7 line (left panel)
and for the 1v5 line (central panel). The 1v7,1f levels are
: 0.5, 0.9, 1.9, 2.9, 3.9 Jy/beam ($3\sigma \approx 0.4$)
and the 1v5,1f levels are in steps of 0.02 Jy/beam starting 
in 0.04 Jy ($\approx 3 \sigma$). Negative levels correspond to 
$2 \sigma$. The cross indicates the position of
the 3mm continuum peak. Right Panel: 
Spectra towards positions T$_{\rm max}$ ($1.2''$ SW of HC1) and HC2. 
T$_{\rm max}$ is $1.8''$ south of source I. The
intensity scale for each spectra is shown in the plot in units of Jy/beam.
\label{f:maps}}
\end{figure}

All the observed vibrationally excited lines have been detected.
Figs. \ref{f:maps}a and \ref{f:maps}b show the spatial distribution
of the HC$_3$N J=10--9 1v7,1f and 1v5,1f transitions (hereafter 1v7 
and 1v5 lines respectively) towards Orion KL.  
The maps represent the integrated intensity of the 1v7 and 
1v5 lines for the main gaussian line (velocity interval from 
3.5 to 7.7 kms$^{-1}$) where the 1v5 line is detected. The most remarkable result is
the different locations of the 3 mm continuum peak 
and those of the 1v7 and of the 1v5 lines. 
While the 1v7 line peaks $1.1''$ NE of the 3 mm peak
(a cross in Fig. 1), the 1v5 line peaks $1.3''$ south of it. 
Since both lines and the continuum were observed simultaneously 
and in the same receiver band, the separations between 
the 1v7, 1v5 and the continuum peaks are not due to instrumental effects. 
From the signal to noise ratio ($\geq11$) the uncertainty in 
the relative positions will be smaller than $0.2''\times0.3''$ 
\citep{plambeck95}. \citet{plambeck95} have measured that source I
is located $1''\pm 0.2''$ west of the 3mm continuum peak. 
Since our data do not resolve source I, in the following we will consider that
the uncertainty in the relative position between the 3mm continuum peak and 
I is $\simeq 0.3''$. Then the separation between source I and the 1v5 peak 
is $\simeq 1.8''\pm 0.4''$.




 High angular resolution images in several molecular species show that
the HC breaks into a number of clumps with different radial velocities
and also likely at different distances from source I
(Migenes et al. 1989; Blake et al. 1996; Chandler \& Wood 1997). The 1v7
and the
1v5 peaks can be identified with clumps observed in other molecular species
in the higher resolution images. The 1v7 and 1v5 peaks clearly appear
in the high angular resolution ($1''$) image of the dust emission at 1.3 mm
\citep{blake96} and the CS emission \citep{chandler97}. Our data 
show that the molecular HC is composed of molecular clumps with 
quite different physical conditions.
In the following we will refer to 
HC1 and HC2 as the 1v5 and 1v7 peaks respectively.
The 1v7 emission is extended with a  deconvolved size of $4'' \times 6''$
(PA: 16$^\circ$), in good agreement with the data from \citet{blake96}. On the 
other hand, the emission from the 1v5 line is unresolved with a deconvolved 
size of $\le 3'' \times\le 2''$ (PA: -4$^\circ$).

\begin{figure}
\plotone{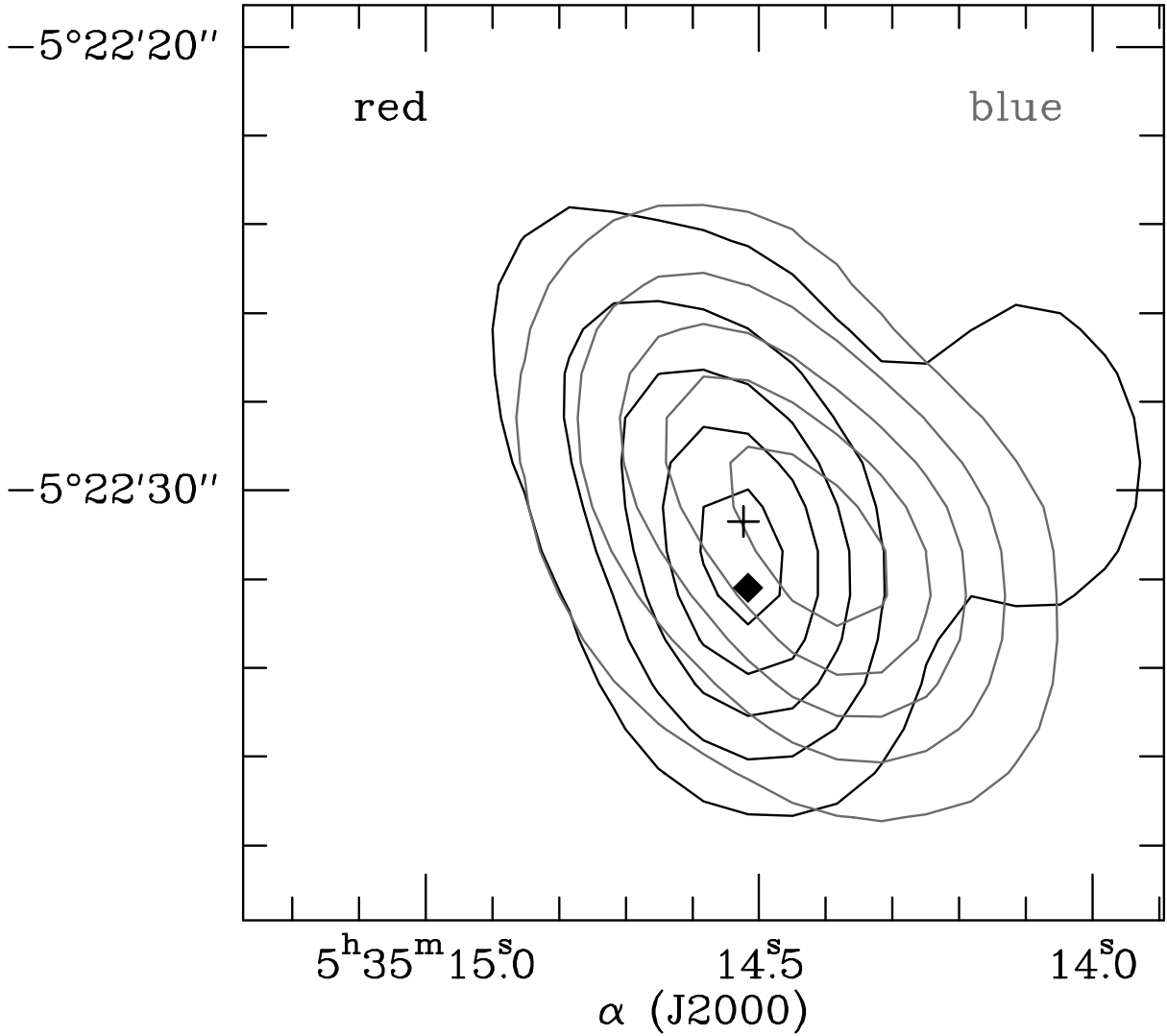}
 \caption{Integrated intensity for the blueshifted and redshifted wings in 
the
HC$_3$N* 1v7,1f line. The blue wing velocity interval is -5 to 1 kms$^{-1}$,
and the red wing is 11 to 17 kms$^{-1}$. 
Levels for the red and blue wings starting in 0.25 Jy/beam 
($\approx 3 \sigma$) in steps of 
0.36 Jy/beam. The cross represents the position for the 3mm continuum, 
and the black diamond the position where the 1v5 line peaks.\label{f:3vels}}
\end{figure}

Fig. \ref{f:maps}c shows the spectra of the 1v7 and the 1v5 lines towards 
HC2 and at T$_{\rm max}$, the  position of maximum temperature where 
the ratio between the 1v5 and 1v7 lines peaks. The 1v5 line shows
a single gaussian
profile centered at a radial velocity of 4.1$\pm$0.5 kms$^{-1}$ and a 
linewidth of 7.0$\pm$0.2 kms$^{-1}$, i.e the typical parameters for the 
Orion KL HC material. The linewidth of the 1v5 line towards T$_{\rm max}$ is
similar to that of the CS clump ``located'' south of source I 
\citep{chandler97}, indicating 
that the excitation of the molecular gas in the HC increases towards the south 
of source I. The 1v7  
lines also show the gaussian profile of the 1v5 line but superimposed on a 
broad pedestal of high-velocity gas. 
Although high-velocity wings in the HC$_3$N emission have 
been reported by \citet{rodriguez92}, this is the first detection
of high-velocity gas in HC$_3$N*. Figure \ref{f:3vels} shows the spatial 
distribution of the high velocity blueshifted (-5 to 1 kms$^{-1}$) 
and redshifted (11 to 17 kms$^{-1}$) 1v7 emission together
with the location of the 3mm peak (cross) and HC1 (filled diamond). 
The high-velocity emission shows a weak bipolar structure around 
I/HC1. The bipolarity is in the same sense
that the large scale molecular outflow observed in CO 
\citep{wilson86}. Very likely, the high-velocity gas detected in 
the 1v7 line is associated with the large bipolar outflow. With our angular 
resolution it is difficult to distinguish the powering source (HC1 or I) 
of the high velocity gas.
As dicussed in section~\ref{excitation} the most likely powering source 
is also the heating source of HC1.

\section{Physical conditions of the molecular gas}

Fig. 1 shows that the ratio between the 1v5 and the 1v7 lines is 
different for HC1, HC2 and the high velocity gas.
One can use the intensities of the vibrationally excited lines 
to estimate the excitation temperature of the molecular gas, using 
the Boltzmann equation. 
Assuming optically thin emission ($\tau \simeq 0.2$ for the 1v7 line, see 
\citet{rodriguez92} for the ground state line), 
we derive excitation temperatures for HC1, HC2 and the high-velocity 
gas of, 230, 170 and 150 K respectively. The excitation 
temperature of 230 K derived for HC1 must be considered a lower 
limit because the contribution from HC2 dominates the 1v7 
emission in HC1. In fact, this excitation temperature is smaller than the
330 K derived from high angular resolution images of the ground state and 
the 1v7 line \citep{wright96}.

The typical error in the excitation temperature is 15\%.
For the excitation temperatures derived above, 
the HC$_3$N column densities for source sizes of $5''$
are  $4\, 10^{15}\ {\rm cm^{-2}}$ and  $8\, 10^{15}\ {\rm cm^{-2}}$
for HC1 and HC2 respectivey, and $6.7\, 10^{14}\ {\rm cm^{-2}}$ and
$2.6\, 10^{15}\ {\rm cm^{-2}}$ for the high velocity gas
(red and blue wings
respectively). 

We can estimate the abundance of HC$_3$N in the hot cores from our
HC$_3$N column density and the H$_2$ column density
obtained by integrating the continuum emission for a source size
of $5''$. Combining our continuum emission at 3.4 mm with the 
integrated continuum emission at 1.3 mm of \citet{blake96} and 
considering a dust temperature of 200 K, similar to that of the HC$_3$N
excitation temperature, we obtain an H$_2$ column density of $2\, 10^{23}\
{\rm cm^{-2}}$ for HC1 and $3\, 10^{23}\ {\rm cm^{-2}}$ for HC2. We obtain an
HC$_3$N abundance of $\simeq 2\, 10^{-8} - 3\, 10^{-8}$ in the hot cores.
The HC$_3$N abundance in the high velocity gas
is more
difficult to derive. This is because the line emission from other molecules
in the velocity range of the HC$_3$N* high velocity gas,
is strongly contaminated by the emission of the 
ambient gas \citep{martin90}.

\section{On the excitation of the HC$_3$N lines} \label{excitation}

 As discussed by \citet{devicente00} and references therein,  
the HC$_3$N* vibrational levels 
are excited by IR emission at wavelengths 
between 14 and  48 $\mu$m. This is because the H$_2$ densities 
of $> 10^{10}\ {\rm cm^{-3}}$ required for the collisional 
excitation 
are at least two orders of magnitude larger than the typical densities 
estimated for the Orion HC. In this case of radiative excitation, the 
excitation temperatures derived from the vibrationally excited lines 
are a good estimate of the dust temperatures within the HC 
because the continuum emission from
HC1 and HC2 at the wavelengths of the vibrational transitions 1v7, 1v6
and 1v5 is optically thick (opacities larger than 10).
For the sizes and the dust temperatures obtained above we derive luminosities 
of $>4\, 10^4$ L$_\odot$ and $4\, 10^4$ L$_\odot$ for HC1 and HC2 respectively. 
  
The luminosity and the location of the possible embedded IR sources
heating HC1 and HC2 can be
better estimated by modeling the excitation and the transfer of the
1v7, 1v6 and 1v5 lines in Orion KL. We have used the model developed
by \citet{devicente00} which considers IR excitation of the vibrational levels 
with a radiation temperature derived from the dust temperature profile 
obtained by \citet{kaufman98} for the Orion HC. In the inner part of the 
HC, the dust is optically thin at the wavelengths of the vibrational 
transitions and the IR radiation temperature differs from the dust temperature.
To account for this effect, we have considered as the inner radius the 
location at which the dust optical depth at the frequency of the 
1v5 transition is equal to unity.

We have considered two cases (parameters are summarized in Table 1): 1)
one IR source located at the position of HC2 and 2) two
IR sources, one located in T$_{\rm max}$ and the other 
in HC2. 

\begin{figure}
\plotone{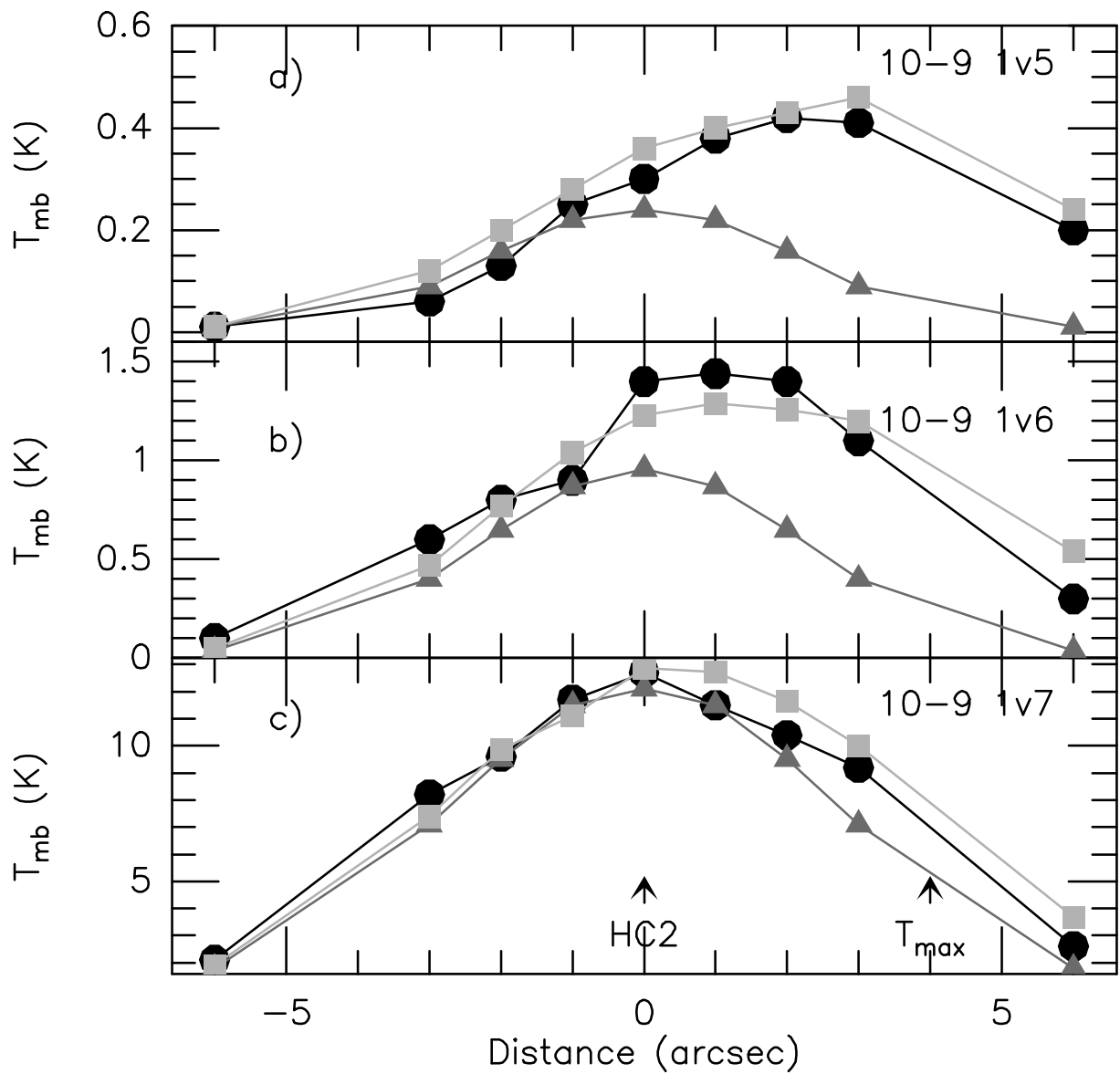}
\caption{Main beam temperature for transitions 10--9,1v7,1f, 10--9,1v6,1f 
and 10--9,1v5,1f for a cut connecting HC2 and HC1 and going from NE to SW. 
Positive offsets are towards SW. The origin of the offset is the 
peak of the 1v7,1f emission. Filled circles indicate the data from the 
PdBI for the channel where the intensity is maximum, triangles are the 
the model predictions for one source, and squares the predicted 
intensities for a model with two sources. Arrows indicate the position of 
the IR sources: T$_{\rm max}$ and HC2. \label{f:dist-model}}
\end{figure}

Figs. \ref{f:dist-model}a, \ref{f:dist-model}b and \ref{f:dist-model}c 
show, respectively,
the spatial distribution of the emission predicted by the model 
for the 1v5, 1v6 and 1v7 lines in a NE-SW cut for one IR source 
(filled triangles) 
and two IR sources (filled squares) superposed on the observed 
intensities (filled circles).
The main conclusion from Fig. \ref{f:dist-model} is that an IR source 
with a moderate luminosity of $2.5\, 10^4$ L$_\odot$ could only fit 
the line intensity and spatial distribution of the 1v7 line 
(panel b).

The 1v5 and 1v6 line distributions requires the presence of a 
another source located close to HC1.
One can reasonably fit the 1v5 and 1v6 lines by considering a second
IR source located $1.5''$ SW of HC1 with a total  
luminosity of $10^5$ L$_\odot$ but with a lower column density of 
hot material than in HC2 (see fig. 3a). This is in agreement with the 
results of \citet{kaufman98} for the proposed scenario of internal heating.
In summary, the HC$_3$N* emission shows that one of the most luminous 
embedded IR sources
($\simeq 10^5$ L$_\odot$) in the Orion KL region must be located in 
the HC $\simeq 1.5''$ SW of HC1; ie. $2.5''$ south of source I.

Since IR radiation is also required for the excitation of the 
HC$_3$N* bipolar outflow, the high velocity material
must be located very close and likely driven by 
the most luminous source in the HC, i.e HC1. 
In fact, the HC$_3$N* 
high velocity gas would trace the material just being accelerated in the 
region where the molecular outflow originates.
The picture emerging from all the HC$_3$N* data 
is that the main heating source, located 
close to HC1, should also drive the molecular bipolar outflow.

\section{Discussion}

The origin of the heating of the HC and of the driving source 
of the molecular outflow in Orion KL has been the subject of debate during 
the last 10 years and it is not settled yet. IRc2 and later source I 
have been the most likely candidates. \citet{kaufman98} have studied the 
heating of the HC by source I and conclude that the HC is internally heated 
by young stars. The detection of the largest excitation temperature derived 
from the 1v5/1v7 ratio, located $\sim 2''$ south of source I, also 
suggests that this source might not be the main heating source. One could 
consider that the excitation temperature gradient can be due to projection 
effects and that HC1 is closer to source I than HC2. In this case, the 
minimum separation between
the 1v5/1v7 maximum and source I would be the projected distance; ie
$\simeq$ 1000 AU. In order to externally heat the H$_2$ column densities 
of HC1 of $\sim 2\, 10^{23} {\rm cm^{-2}}$ up to temperatures of 230 K, source 
I should have a luminosity of $> 5\, 10^{5} L_\sun$ \citep{kaufman98}.
This is larger than the total luminosity of the KL region \citep{werner76}.
Heating of HC1 by the combination of sources I and n will not substantially
change the required total luminosity of source I since 
source n is considered to be less luminous than source I and is located
at the same projected distance from HC1 as source I. IR radiation from the 
shocks generated in the molecular outflows can be ruled out since the mechanical
luminosity of the outflow is only $10^2\ L_\sun$ \citep{rodriguez99}. We 
conclude that the most likely explation for the heating of HC1 is internal 
heating by young embedded stars. The lack of radio emission from HC2 
might be an indication of mass accretion \citep{walmsley95}.


The high luminosity ($\sim 10^5$ L$_\odot$) IR source
heating HC1 is, like source I, a very weak emitter at 20$\mu$m.
The 1v5 line emission is
anticorrelated with the IR emission at 20$\mu$m measured by \citet{gezari98}
with $1''$ resolution.
The morphology of the IR emission is possibly influenced 
by foreground extinction \citep{wright96}, and HC1 can be obscured by the
presence of foreground cold material. From the parameters in Table 1 for 
T$_{max}$, the expected flux density within $1''$ at 20$\mu$m from HC1  
should be $4\, 10^{3}$ Jy, a factor of 
$\approx 500$ larger than the observed flux. To match the observed 
and predicted fluxes within $1''$, the IR source in HC1 must be
hidden behind 6 magnitudes of extinction at 20 $\mu$m, i.e. an H$_2$ 
column density of $1.2\, 10^{23}\, {\rm cm^{-2}}$.
As expected, the cold gas in front of the IR source not observed in the
HC$_3$N* lines is 20\% of the HC1 total column
density of $4\, 10^{23}\, {\rm cm^{-2}}$, derived from the dust continuum 
emission \citep{blake96} also measured with $1''$ resolution. 
Then, the IR map at 20 $\mu$m seems to be consistent with our estimates 
for the luminosity of the IR source embedded in HC1.

Source I is the driving source of the SiO outflow, but is located
just outside the HC. Furthermore, the molecular outflow
observed in HC$_3$N*  seems to be driven by the IR source in HC1.
These facts suggest, as
proposed by \citet{rodriguez99}, that the Orion KL region contains multiple
molecular outflows.

The data presented in
this paper, the presence of a high luminosity IR source
embedded in HC1 powering the high-velocity gas observed
in HC$_3$N* emission and the properties of the IR emission
in the Orion KL region (morphology, luminosity and polarization) can be
explained in the scenario of a young massive star deepely embbeded 
in molecular material \citep{plambeck82}. The circumstellar material 
must be very anisotropic since it is located in the narrow ridge of 
minimum IR emission with strong molecular line emission
and has large column densities along the line of sight.
Although our angular resolution is not enough to resolve the circumstellar
material, the most likely morphology is that of an edge-on disk.
The circumstellar material is
internally heated by young massive protostars with a
luminosity of $10^5$ L$_\odot$. The IR radiation heats
the most inner parts giving rise to the
emission observed in HC$_3$N*.
The IR radiation escapes in the direction of the lowest
column densities (perperdicular to the ridge)
illuminating most of the IR sources
as shown by the polarization vectors which point towards a
location in the vicinity of HC1 \citep{werner83}. The massive protostar
drives the 
large molecular outflow observed in CO that is collimated by the circumstellar
material. The material very close to the IR source
where the mass ejection originates is observed as high velocity wings in
HC$_3$N*.

This work has been partially supported by the Spanish Ministerio de Ciencia 
y Tecnologia grant ESP-4519-PE and SEPCT grant AYA2000-927.
We want to thank the anonymous referees for their constructive comments which
helped to improve substantially the original version of this letter.

\clearpage

\clearpage

\begin{table}
 \caption{Final parameters used in the model.}
 \begin{flushleft}
 \begin{tabular}{lrr}
 \multicolumn{1}{l}{Parameter} & \multicolumn{1}{c}{HC2} 
 & \multicolumn{1}{c}{T$_{\rm max}$} \\ \hline
 Luminosity (L$_\odot$)           &   $2.5\, 10^4$ &  $10^5$ \\
 n(H$_2$) (cm$^{-3}$, $r_0=10^{16} $cm)
 &$7\, 10^7 (r/r_0)^{-3/2}$                  & $5\, 10^7 (r/r_0)^{-3/2}$ \\
 HC$_3$N line width (kms$^{-1}$)    &         6     &  6 \\
 HC$_3$N abundance                  &$ 7\, 10^{-9}$ & $8\, 10^{-9}$  \\
 Source size depth (arcsec)         &         15    &  4  \\
 Source size width (arcsec)         &          4    &  4  \\
 HPBW (arcsec)                      &          4    &  4  \\
 Offset of source 1 from 2 (arcsec) &               &  4  \\
 \hline
 \end{tabular}
 \end{flushleft}
 \label{t:model}
\end{table}

\end{document}